\documentclass[preprint,12pt]{elsarticle}
\usepackage{graphicx}
\usepackage{amssymb}
\usepackage{amsmath}
\usepackage{subfig}
\usepackage{multirow}
\usepackage{colortbl,color}
\usepackage{algorithm}
\usepackage{amsthm}
\usepackage{hyperref}
\biboptions{comma,sort&compress}
\journal{ }

\begin{document}

\begin{frontmatter}
\title{Improved Diffuse Boundary Condition for the DSBGK Method to Eliminate the Unphysical Density Drift}
\author[KAUST]{Jun~Li\footnote[1]{e-mail: lijun04@gmail.com \\ To view the network videos of the DSBGK transient results of several benchmark problems, please click the \href{https://www.dropbox.com/sh/w6t5s7cyteo60fs/Nb5umVXBIU}{\textit{Dropbox link}} or find the link in \href{http://dl.dropboxusercontent.com/u/13266492/JunLiCV.pdf}{\textit{CV}}.}}
\address[KAUST]{Applied Mathematics and Computational Science\\ King Abdullah University of Science and Technology\\ Thuwal, Saudi Arabia}
\begin{abstract}
An improved diffuse boundary condition, where the number flux of the incoming real molecules on the wall surface is calculated using the molecular variables rather than the cell's macroscopic variables, is proposed to eliminate the unphysical density drift, which was observed in the previous DSBGK simulation of the lid-driven problem but disappears in the channel flow problem because of the density constraint imposed at the open boundaries. Consequently, the efficient time-average process is valid for sampling all quantities of interest in closed as well as open problems. When the driven velocity of the lid-driven problem is only one micrometer per second that is realistic in the micro-electro-mechanical systems or pore-scale flows of the shale gas, we suggest to use the original boundary condition because the density drift during a very long time interval becomes unperceivable when the perturbation is tiny. Thus, the original diffuse boundary condition, which contains much less stochastic noise than the improved one, is still an advisable choice in closed problems with tiny perturbation and in open problems unless the density drift occurs and fails the goal of the simulations making the use of the improved boundary condition necessary. 
\end{abstract}
\end{frontmatter}
\section{Introduction}\label{s:intro}
For micro gas flows, the Boltzmann equation rather than the Navier-Stokes equation should be used due to the high Knudsen number $Kn=\lambda/L$ where $\lambda$ is the molecular mean free path and $L$ is the characteristic length of the flow problem. The characteristic velocity of micro gas flows is usually much smaller than the molecular random thermal velocity and sometimes the variations of quantities of interest inside the flow domain are very small, which makes the computational cost of the traditional DSMC method \cite{Graeme1994} prohibitive although it is efficient in the case of high-speed flows. The design of the micro-electro-mechanical systems (MEMS) promoted the development of numerical simulation methods devoted to the low-speed problems of high $Kn$. Recently, the simulations of pore-scale flows of the shale gas, where the pore size is about 10-100 nm and the flow velocity is tiny, are desirable since the experimental measures to get the permeability of representative rock samples may take a couple of months. 

The DSBGK method, which is proposed in \cite{Jun2011RGD} and detailed in \cite{Jun2012arXiv}, is very efficient for simulating low-speed problems of high $Kn$ and verified against the DSMC method in the lid-driven, Couette, channel flow and thermal transpiration problems over a wide range of $Kn$ \cite{Jun2012arXiv}. In addition to its high-efficiency, the DSBGK method has many numerical advantages including simplicity, stability, convenience for complex configuration and for parallel computation because the basic algorithmic structure of the DSMC method is employed.

In the original diffuse boundary condition, we use the cell's variables to compute the number flux of incoming real molecules on the wall surface such that the transient results contain low stochastic noise. This also reduces the memory usage in open problems where the DSBGK simulation remains stable when using only $10$ simulated molecules per cell \cite{Jun2012arXiv}. But, in closed problems, the density drift occurs due to the lack of density constraint that is imposed at the open boundaries of open problems. Although the DSBGK simulations of closed problems using about $10$ simulated molecules per cell are also stable (see Fig.~\ref{fig:CavityflowKn6.3-smallU}), the density drift prohibits the application of the time-average process used to reduce the stochastic noise which is obvious when using such small amount of simulated molecules. We propose an improved boundary condition here to eliminate the unphysical density drift and then the efficient time-average process is valid for sampling all quantities of interest in closed as well as open problems. The DSBKG steady state results obtained using the time-average process are verified in the lid-driven problem as an example. 
\section{DSBGK Method}\label{s:DSBGK method}
We consider the gas flows of single component. In the absence of external body force, the BGK equation \cite{BGK1954} can be written as a Lagrangian form:
\begin{equation}\label{eq:BGK}
    \dfrac{\mathrm{d}f}{\mathrm{d}t}=\dfrac{\partial f}{\partial t}+c_j\dfrac{\partial f}{\partial x_j}=\upsilon(f_{\mathrm{eq}}-f),
\end{equation}
where $f(t, \vec x, \vec c)$ is the unknown probability distribution function, $t$ the time, $\vec x$ the spatial coordinate, $\vec c$ the molecular velocity and, the parameter $\upsilon$ is selected appropriately to satisfy the coefficient of dynamic viscosity $\mu$ or heat conduction $\kappa$ as follows \cite{Vincenti1965}:
\begin{equation}\label{eq:upsilon}
    \begin{cases}
    \mu_{\mathrm{BGK}}=\dfrac{nk_\mathrm{B}T}{\upsilon} \\
    \kappa_{\mathrm{BGK}}=\dfrac{5k_\mathrm{B}}{2m}\dfrac{nk_\mathrm{B}T}{\upsilon},
    \end{cases}
\end{equation}
and the Maxwell distribution function $f_{\mathrm{eq}}$ is:
\begin{equation}\label{eq:feq}
    f_{\mathrm{eq}}(t, \vec x, \vec c)=n(\dfrac{m}{2\pi k_\mathrm{B}T})^{3/2}\exp[\dfrac{-m(\vec c-\vec u)^2}{2k_\mathrm{B}T}],
\end{equation}
where $m$ is the molecular mass, $k_\mathrm{B}$ the Boltzmann constant and, the number density $n$, flow velocity $\vec u=(u_x, u_y, u_z)$ and temperature $T$ are functions of $t$ and $\vec x$ and defined using the integrals of $f$ with respect to $\vec c$.

The DSBGK method is proposed in \cite{Jun2011RGD} and detailed in \cite{Jun2012arXiv} where the extension to problems with external force is discussed. The simulation process is divided into a series of time steps $\Delta t$ and the computational domain is divided into many cells. $\Delta t$ and the cell size $\Delta x$ are selected the same as in the DSMC method when simulating problems of high $Kn$. Each simulated molecule $l$ carries four molecular variables: position $\vec x_l$, molecular velocity $\vec c_l$,  number $N_l$ of real molecules represented by the simulated molecule $l$, and $F_l$ which is equal to $f(t, \vec x_l, \vec c_l)$. The variables $n_{\mathrm{tr,}k}, \vec u_{\mathrm{tr,}k}, T_{\mathrm{tr,}k}$ of each cell $k$ are updated using $\vec x_l, \vec c_l$ and the increment of $N_l$ based on the mass, momentum and energy conservation principles of the intermolecular collision process and are simultaneously used to update the molecular variables based on the BGK equation and the extrapolation \cite{Jun2012arXiv} of acceptance-rejection scheme. The DSBGK method is a molecular simulation method and theoretically all macroscopic quantities of interest should be computed using the molecular variables. The cell's varialbes $n_{\mathrm{tr,}k}, \vec u_{\mathrm{tr,}k}, T_{\mathrm{tr,}k}$ are auxiliary variables, which replace the original macroscopic quantities $n_k, \vec u_k, T_k$ defined by the molecular variables to reduce the stochastic noise, and converge to $n_k, \vec u_k, T_k$ as discussed after Eq.~(13) of \cite{Jun2012arXiv}.

\subsection{Summary of the DSBGK algorithm}\label{ss:algorithm summary}
1. Initialization. Generate many cells and simulated molecules and assign them initial values for $n_{\mathrm{tr,}k}, \vec u_{\mathrm{tr,}k}, T_{\mathrm{tr,}k}$ and $\vec x_l, \vec c_l, F_l, N_l$, respectively.

2. Each simulated molecule moves uniformly and in a straight line before encountering boundaries. During each $\Delta t$, the trajectory of each particular molecule $l$ may be divided into several segments by the cell's interfaces. Then, $\vec x_l, F_l, N_l$ are \textit{deterministically} updated along each segment in sequence. When encountering the wall boundaries, $\vec c_l$ is randomly updated according to the reflection model and then $F_l$ is updated correspondingly. In open problems, simulated molecules are removed from the computational domain when moving across the open boundaries during each $\Delta t$ and new simulated molecules are generated at the open boundaries after each $\Delta t$. The variables $n_{\mathrm{tr,}k}, \vec u_{\mathrm{tr,}k}, T_{\mathrm{tr,}k}$ of each cell $k$ are updated after each $\Delta t$.

3. After convergence, $n_{\mathrm{tr,}k}, \vec u_{\mathrm{tr,}k}, T_{\mathrm{tr,}k}$ are used as the discrete solutions of the BGK equation at steady state.

\subsection{Wall boundary conditions}\label{ss:wall boudary condition}
In the reflection models of wall boundary, $\vec c_l$ and then $F_l$ are changed after molecular reflection at $\vec x_l$ on the wall. $N_l$ remains unchanged to conserve the mass. The reflecting velocity $\vec c_\mathrm{r}$ is randomly selected the same as in the DSMC method and then $\vec c_l$ is updated to $\vec c_{\rm r}+\vec u_{\rm wall}$ where $\vec u_{\rm wall}$ is the wall velocity (see the details in \cite{Jun2012arXiv}). The subscript $l$ is omitted in the notations of the incoming and reflecting velocities.

As discussed in \cite{Jun2012arXiv}, $F_l$ is updated to $F_{l\mathrm{,new}}=f(t, \vec x_l, \vec c_{l\mathrm{,new}})=f(t, \vec x_l, \vec c_\mathrm{r}+\vec u_{\mathrm{wall}})$ after getting $\vec c_\mathrm{r}$. Note that $F_l$ is the representative value of $f$ which is different from the scatter kernel $R$ used to select $\vec c_\mathrm{r}$ for each particular reflection process. Generally speaking, $f$ is related to the mass flux but $R$ has nothing to do with the mass flux. Usually, $R$ describes the probability distribution of $\vec c_\mathrm{r}$ inside the half velocity space ($\vec c_\mathrm{r}\cdot\vec e_\mathrm{n}>0$ where $\vec e_\mathrm{n}$ is the outer-normal unit vector of the wall) as a function depending on the wall temperature $T_\mathrm{wall}$, accommodation coefficients $\alpha_\mathrm{n}, \alpha_\tau$ and possibly also on the incoming velocity $\vec c_\mathrm{i}$. So, we have $R=R(\vec c_\mathrm{r}, \vec c_\mathrm{i})$ which contains $T_\mathrm{wall}, \alpha_\mathrm{n}, \alpha_\tau$ as parameters. $R$ satisfies the normalization condition $\int_{\vec c_\mathrm{r}\cdot\vec e_\mathrm{n}>0}R(\vec c_\mathrm{r}, \vec c_\mathrm{i})\mathrm{d}\vec c_\mathrm{r}=1$ where $R(\vec c_\mathrm{r}, \vec c_\mathrm{i})\mathrm{d}\vec c_\mathrm{r}$ is the probability for the molecule coming at $\vec c_\mathrm{i}$ to reflect into the velocity space element $\mathrm{d}\vec c_\mathrm{r}$ around $\vec c_\mathrm{r}$. The transformation between $f$ at the wall boundary and $R$ can be completed using the incoming mass flux.

We introduce $f_\mathrm{B}(\vec c)$ as the equivalent distribution function of $f$ observed at the reflection point $\vec x_l$ and at the current moment $t$ in a local Cartesian reference frame $S_\mathrm{local}$ moving at $\vec u_{\rm wall}$, which means $f_\mathrm{B}(\vec c)=f(t, \vec x_l, \vec c+\vec u_\mathrm{wall})$. After getting the formula of $f_\mathrm{B}(\vec c)$, we have $F_{l\mathrm{,new}}=f_\mathrm{B}(\vec c_\mathrm{r})$. The distribution $f_\mathrm{B}(\vec c_\mathrm{i})|_{\vec c_\mathrm{i}\cdot\vec e_\mathrm{n}<0}$ of the incoming molecules is known from the molecular information in the adjacent cell. $f_\mathrm{B}(\vec c_\mathrm{r})|_{\vec c_\mathrm{r}\cdot\vec e_\mathrm{n}>0}$ is the distribution of reflecting molecules and related to $R$ as introduced in \cite{Ching2005}:
\begin{equation}\label{eq:kernel}
    f_\mathrm{B}(\vec c_\mathrm{r})(\vec c_\mathrm{r}\cdot\vec e_\mathrm{n})\mathrm{d}\vec c_\mathrm{r}=-\int_{\vec c_\mathrm{i}\cdot\vec e_\mathrm{n}<0}R(\vec c_\mathrm{r}, \vec c_\mathrm{i})f_\mathrm{B}(\vec c_\mathrm{i})(\vec c_\mathrm{i}\cdot\vec e_\mathrm{n})\mathrm{d}\vec c_\mathrm{i}\mathrm{d}\vec c_\mathrm{r}.
\end{equation}
Taking integration of Eq.~\eqref{eq:kernel} with respect to $\vec c_\mathrm{r}$ over its half velocity space and using the normalization condition of $R(\vec c_\mathrm{r}, \vec c_\mathrm{i})$, we get:
\begin{equation}\label{eq:massbalance}
\begin{aligned}
    &\int_{\vec c_\mathrm{r}\cdot\vec e_\mathrm{n}>0}f_\mathrm{B}(\vec c_\mathrm{r})(\vec c_\mathrm{r}\cdot\vec e_\mathrm{n})\mathrm{d}\vec c_\mathrm{r} \\
    &=-\int_{\vec c_\mathrm{r}\cdot\vec e_\mathrm{n}>0}\int_{\vec c_\mathrm{i}\cdot\vec e_\mathrm{n}<0}R(\vec c_\mathrm{r}, \vec c_\mathrm{i})f_\mathrm{B}(\vec c_\mathrm{i})(\vec c_\mathrm{i}\cdot\vec e_\mathrm{n})\mathrm{d}\vec c_\mathrm{i}\mathrm{d}\vec c_\mathrm{r} \\
    &=-\int_{\vec c_\mathrm{i}\cdot\vec e_\mathrm{n}<0}f_\mathrm{B}(\vec c_\mathrm{i})(\vec c_\mathrm{i}\cdot\vec e_\mathrm{n})\mathrm{d}\vec c_\mathrm{i},
\end{aligned}
\end{equation}
which represents the mass conservation of molecular reflection process.

In the Maxwell diffuse reflection model, we have:
\begin{equation}\label{eq:f_B-cr-Maxwell}
\begin{aligned}
    f_\mathrm{B,diffuse}(\vec c_\mathrm{r})=n_\mathrm{eff}(\dfrac{m}{2\pi k_\mathrm{B}T_\mathrm{wall}})^{3/2}\exp(\dfrac{-m\vec c_\mathrm{r}^2}{2k_\mathrm{B}T_\mathrm{wall}}),
\end{aligned}
\end{equation}
where the effective $n_\mathrm{eff}$ is determined from $f_\mathrm{B}(\vec c_\mathrm{i})$ based on the mass conservation principle of Eq.~\eqref{eq:massbalance}. Theoretically, $f_\mathrm{B}(\vec c_\mathrm{i})$ depends on the incoming molecules. To reduce the stochastic noise, we previously use the cell's variables rather than the molecular variables to determine $f_\mathrm{B}(\vec c_\mathrm{i})$ as follows:
\begin{equation}\label{eq:f_B-ci-bad}
\begin{aligned}
    f_\mathrm{B,bad}(\vec c_\mathrm{i})=n_{\mathrm{tr,}k}(\dfrac{m}{2\pi k_\mathrm{B}T_{\mathrm{tr,}k}})^{3/2}\exp[\dfrac{-m(\vec c_\mathrm{i}-(\vec u_{\mathrm{tr,}k}-\vec u_\mathrm{wall}))^2}{2k_\mathrm{B}T_{\mathrm{tr,}k}}],
\end{aligned}
\end{equation}
where $n_{\mathrm{tr,}k}, \vec u_{\mathrm{tr,}k}, T_{\mathrm{tr,}k}$ are the quantities of cell $k$ close to the reflection point $\vec x_l$. Then, the number of incoming real molecules on per unit wall area during per unit time is:
\begin{equation}\label{eq:Num-in-bad}
\begin{aligned}
    N_\mathrm{in,bad}&=-\int_{\vec c_\mathrm{i}\cdot\vec e_\mathrm{n}<0}f_\mathrm{B,bad}(\vec c_\mathrm{i})(\vec c_\mathrm{i}\cdot\vec e_\mathrm{n})\mathrm{d}\vec c_\mathrm{i} \\
    &=n_{\mathrm{tr,}k}\sqrt{\dfrac{k_\mathrm{B}T_{\mathrm{tr,}k}}{2\pi m}}[\exp(-{u'}_\mathrm{in}^2)+\sqrt{\pi}{u'}_\mathrm{in}(1+\mathrm{erf}({u'}_\mathrm{in}))],
\end{aligned}
\end{equation}
where ${u'}_\mathrm{in}=\dfrac{-(\vec u_{\mathrm{tr,}k}-\vec u_\mathrm{wall})\cdot\vec e_\mathrm{n}}{\sqrt{2k_\mathrm{B}T_{\mathrm{tr,}k}/m}}$. Similarly, the number $N_\mathrm{out}$ of reflecting real molecules is:
\begin{equation}\label{eq:Num-out-Maxwell}
\begin{aligned}
    N_\mathrm{out}=\int_{\vec c_\mathrm{r}\cdot\vec e_\mathrm{n}>0}f_\mathrm{B,diffuse}(\vec c_\mathrm{r})(\vec c_\mathrm{r}\cdot\vec e_\mathrm{n})\mathrm{d}\vec c_\mathrm{r}
    =n_\mathrm{eff}\sqrt{\dfrac{k_\mathrm{B}T_\mathrm{wall}}{2\pi m}}.
\end{aligned}
\end{equation}
Let $N_\mathrm{out}=N_\mathrm{in,bad}$ as required by Eq. \eqref{eq:massbalance}, we get an estimation of $n_\mathrm{eff}$:
\begin{equation}\label{eq:n-eff-Maxwell-bad}
\begin{aligned}
    n_\mathrm{eff,bad}=n_{\mathrm{tr,}k}\sqrt{\dfrac{T_{\mathrm{tr,}k}}{T_\mathrm{wall}}}[\exp(-{u'}_\mathrm{in}^2)+\sqrt{\pi}{u'}_\mathrm{in}
    (1+\mathrm{erf}({u'}_\mathrm{in}))].
\end{aligned}
\end{equation}
After getting $n_\mathrm{eff,bad}$, we can determine $F_{l\mathrm{,new}}=f_\mathrm{B,diffuse}(\vec c_\mathrm{r})$ where $n_\mathrm{eff}=n_\mathrm{eff,bad}$. We store $n_\mathrm{eff,bad}$ and repeatedly use it for different simulated molecules reflecting on the same subarea adjacent to the cell $k$ during the same $\Delta t$ and update $n_\mathrm{eff,bad}$ after each $\Delta t$.

The applications of the CLL reflection model \cite{Carlo1971CLL}-\cite{Lord1991CLL} and the specular reflection model are discussed in \cite{Jun2012arXiv}.

\section{Improved Diffuse Reflection Model}\label{s:improved BCs}
To update $F_l$ after reflecting on the wall, the diffuse and CLL reflection models need the distribution $f_\mathrm{B}(\vec c_\mathrm{i})$ of all incoming molecules but the specular reflection model only uses the information of each particular incoming molecule. In the above diffuse reflection model, we use Eq.~\eqref{eq:f_B-ci-bad} as $f_\mathrm{B}(\vec c_\mathrm{i})$ for simplicity to calculate the incoming number flux such that the stochastic noise is reduced. This simplification is reasonable when $f_\mathrm{B}(\vec c)$ near the boundary is close to the local Maxwell distribution where the incoming number flux depends only on $n, \vec u$ and $T$. But, the deviation of $f_\mathrm{B}(\vec c)$ from the Maxwell distribution usually originates from the boundary and could be large at the boundary. The accuracy of Eq.~\eqref{eq:f_B-ci-bad} depends on the magnitude of deviation. 

Actually, the incoming number flux can be directly calculated using the information of incoming molecules although the correspondingly computed flux contains high stochastic noise since the number of simulated molecules per cell is usually small. As in the DSMC method, it is convenient for the DSBGK method to calculate the \textit{net} flux $\Gamma(Q)$ of any molecular quantity $Q(\vec c)$ in unit time and across unit area of the boundary surface:
\begin{equation}\label{eq:net-flux}
    \Gamma(Q)=\dfrac{1}{\Delta t\Delta S}\sum_lN_l[Q(\vec c_\mathrm{i})-Q(\vec c_\mathrm{r})]_l,
\end{equation}
where the summation is over the simulated molecules reflecting on the subarea $\Delta S$ during the current time step $\Delta t$, $Q(\vec c_\mathrm{i})$ and $Q(\vec c_\mathrm{r})$ are the incoming and reflecting quantities, respectively. Let $Q=m\vec c$ and $m\vec c^2/2$ and then $\Gamma(Q)$ represents the stress and heat flux, respectively. Similarly, the incoming number flux is computed as:
\begin{equation}\label{eq:Num-in-actual}
    N_{\rm in}=\dfrac{1}{\Delta t\Delta S}\sum_lN_l.
\end{equation}
Let $N_\mathrm{out}=N_\mathrm{in}$ as required by Eq.~\eqref{eq:massbalance}, we get an improved formula for $n_\mathrm{eff}$:
\begin{equation}\label{eq:n-eff-Maxwell-actual}
\begin{aligned}
    n_\mathrm{eff}=\sqrt{\dfrac{2\pi m}{k_\mathrm{B}T_\mathrm{wall}}}\dfrac{1}{\Delta t\Delta S}\sum_lN_l.
\end{aligned}
\end{equation}
$F_{l\mathrm{,new}}=f_\mathrm{B,diffuse}(\vec c_\mathrm{r})$ is implemented to update $F_l$ using Eq.~\eqref{eq:f_B-cr-Maxwell} during the simulation process and $n_\mathrm{eff}$ is updated using Eq.~\eqref{eq:n-eff-Maxwell-actual} after each $\Delta t$. The division of $\Delta S$ is based on the cell's division and each of those cell's interfaces located at the wall surfaces has a variable $n_{\rm eff}$. At the initial state with Maxwell distribution and $(\vec u_{{\rm tr},k}-\vec u_{\rm wall})\cdot\vec e_{\rm n}=0$, we have $n_\mathrm{eff}|_{t=0}=n_{{\rm tr},k}\sqrt{\dfrac{T_{\mathrm{tr,}k}}{T_\mathrm{wall}}}$. We note that $N_{\rm in,bad}$ is of the same order of magnitude as the expected value of $N_{\rm in}$ since the incoming number flux depends mostly on the macroscopic quantities, which accunts for the favourable simulation results obtained using $N_{\rm in,bad}$ \cite{Jun2012arXiv}. Although the use of the improved boundary condition leads to negligible increase of the computational time for each time step, more samples are required to reduce the stochastic noise of average results because $N_{\rm in}$ contains more stochastic noise than $N_{\rm in, bad}$ as implied in Fig.~\ref{fig:evolution of total number}.

In addition to the small improvement of accuracy (see the above discussion and compare Fig.~\ref{fig:CavityflowKn6.3U20-time-average} with Fig. 5 of \cite{Jun2012arXiv}), another advantage (see Fig.~\ref{fig:evolution of total number}) of using the molecular variables to compute the incoming number flux by Eq.~\eqref{eq:Num-in-actual} is to eliminate the unphysical drift of $n_{{\rm tr},k}$, which was observed in the previous simulations of \textit{closed} problems using $N_{\rm in, bad}$ and disappears in the open problems because of the fixed density imposed at the open boundaries \cite{Jun2012arXiv}. We didn't observe drift in the distributions of $\vec u_{{\rm tr},k}$ and $T_{{\rm tr},k}$, which implies that the velocity $\vec u_k$ and temperature $T_k$ defined using the molecular variables as in Eq.~(3) of \cite{Jun2012arXiv} are averagely equal to $\vec u_{{\rm tr},k}$ and $T_{{\rm tr},k}$, respectively (see the discussion after Eq.~(13) of \cite{Jun2012arXiv}). In order to simplify the analysis, we assume that $\vec u_{{\rm tr},k}=\vec u_{\rm wall}$, $T_{{\rm tr},k}=T_{\rm wall}$ and the distribution of simulated molecules is the local Maxwell distribution in each cell $k$ adjacent to the wall. Thus, $N_{\rm in}$ computed by Eq.~\eqref{eq:Num-in-actual} is averagely equal to $n_k\sqrt{\dfrac{k_\mathrm{B}T_k}{2\pi m}}$ and $n_\mathrm{eff}$ computed by Eq.~\eqref{eq:n-eff-Maxwell-actual} is averagely equal to $n_k=\sum N_l/V_k$ where $\sum N_l$ is the total number of real molecules in the cell $k$ and $V_k$ is the volume of cell $k$. When $n_k<n_{{\rm tr},k}$, $n_\mathrm{eff}$ is smaller than $n_{{\rm tr},k}$ and thus, during the intermolecular collision process in the cell $k$, $F_{l,{\rm new}}=f_\mathrm{B,diffuse}(\vec c_\mathrm{r})$ after reflecting on the wall computed by Eq.~\eqref{eq:f_B-cr-Maxwell} is smaller than $f_{{\rm eq, tr},k}(\vec c_{l,{\rm new}})$ making the corresponding $\Delta_k N_l$ positive, which provides a mechanism to increase $n_k$ and decrease $n_{{\rm tr},k}$ as required by the convergence where $n_k=n_{{\rm tr},k}$ is satisfied on average. In contrast, if the incoming number flux is computed by Eq.~\eqref{eq:Num-in-bad} using the cell's variables, we have $n_\mathrm{eff}=n_\mathrm{eff,bad}=n_{{\rm tr},k}$ according to Eq.~\eqref{eq:n-eff-Maxwell-bad} and then, during the intermolecular collision process in the cell $k$, $F_{l,{\rm new}}$ after reflecting on the wall is equal to $f_{{\rm eq, tr},k}(\vec c_{l,{\rm new}})$ making the previous mechanism of convergence disappeared. Similar analysis applies to the case of $n_k>n_{{\rm tr},k}$. In real applications where the above assumptions for simplicity of analysis are not true, the use of actual incoming number flux computed by Eq.~\eqref{eq:Num-in-actual} also improves the numerical performance with respect to convergence because: 1) $n_\mathrm{eff}$ computed by Eq.~\eqref{eq:n-eff-Maxwell-actual} using the molecular variables is close to $n_k$ on average even though the molecular distribution near the wall deviates from the Maxwell distribution and $\vec u_k=\vec u_{{\rm tr},k}\ne\vec u_{\rm wall}$, $T_k=T_{{\rm tr},k}\ne T_{\rm wall}$; 2) the magnitudes of $F_{l,{\rm new}}=f_\mathrm{B,diffuse}(\vec c_\mathrm{r})$ and $f_{{\rm eq, tr},k}(\vec c_{l,{\rm new}})$ depend mostly on $n_\mathrm{eff}$ and $n_{{\rm tr},k}$, respectively, since $\vec u_{{\rm tr},k}-\vec u_{\rm wall}$ and $T_{{\rm tr},k}-T_{\rm wall}$ are relatively small.

\section{Validate the Improved Diffuse Reflection Model}\label{s:simulations with improved BCs}
In DSBGK simulations, the parameter $\upsilon$ is selected to satisfy the coefficient of dynamic viscosity $\mu$ or heat conduction $\kappa$ by Eq. \eqref{eq:upsilon}. For problems where the momentum exchange is the dominant effect, we use $\upsilon=\upsilon(\mu)\equiv nk_\mathrm{B}T/\mu$ to satisfy $\mu$. For problems where the heat conduction is the dominant effect, we select $\upsilon=\upsilon(\kappa)\equiv5nk_\mathrm{B}^2T/(2m\kappa)$ to satisfy $\kappa$. Note that $\mu$ is given usually. For monoatomic gas where the Prandtl number is $Pr=2/3$ and the specific heat capacity at constant pressure is $C_p=5k_\mathrm{B}/(2m)$, we have $\upsilon(\kappa)=2nk_\mathrm{B}T/(3\mu)$ as $\kappa=C_p\mu/Pr$. Additionally, we define the Knudsen number $Kn=\lambda/L$ of the problem concerned using the initial molecular mean free path $\lambda_0=\dfrac{16\mu}{5n_0\sqrt{2\pi mk_{\rm B}T_0}}$ \cite{Ching2005}, where $n_0$ and $T_0$ are the initial density and temperature, respectively.
\begin{figure}[H] 
\centering
\includegraphics[width=0.35\textwidth]{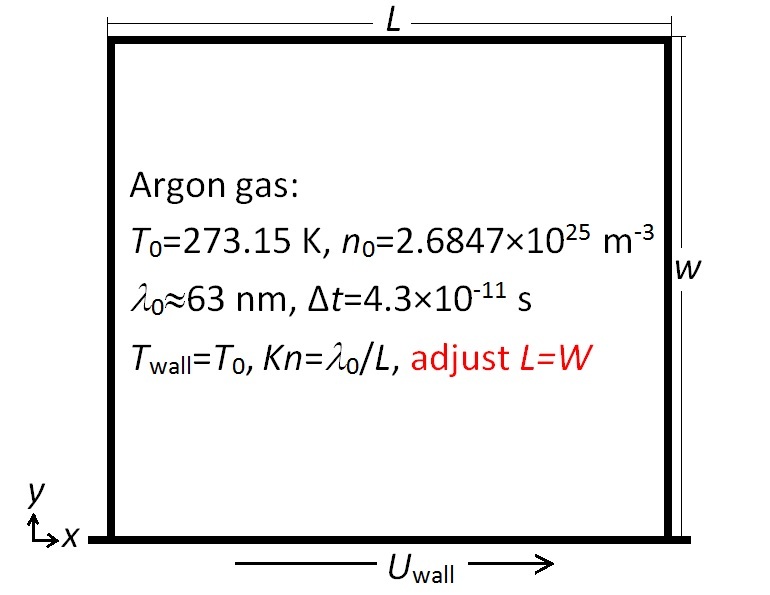}
\caption{Schematic model of the lid-driven cavity flow.}
\label{fig:CavityModel}
\end{figure}

\begin{figure}[H]
  \centering
  \includegraphics[width=0.45\textwidth]{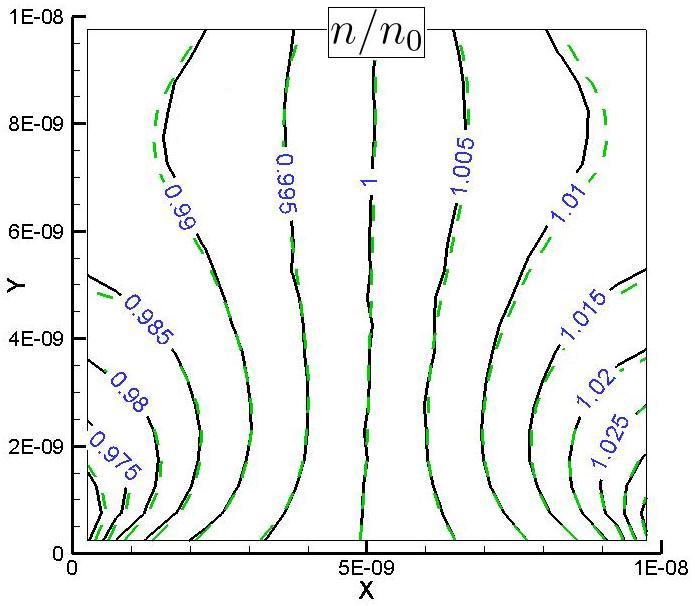}
  \includegraphics[width=0.45\textwidth]{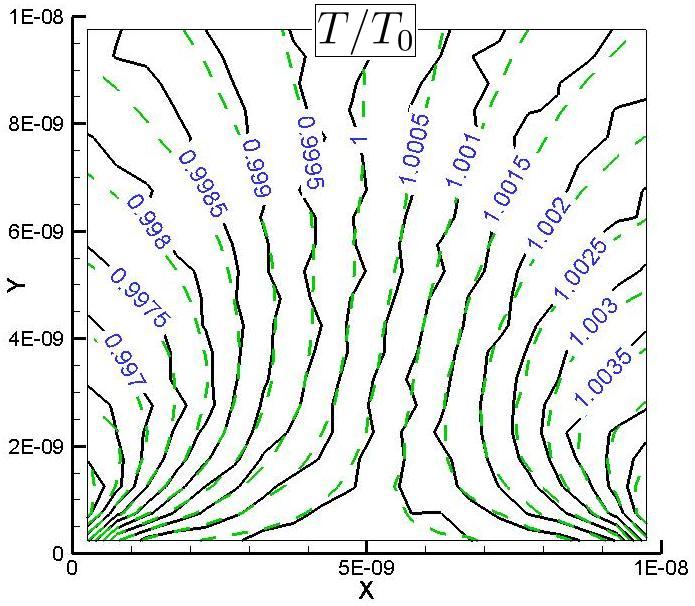} \\
  \includegraphics[width=0.45\textwidth]{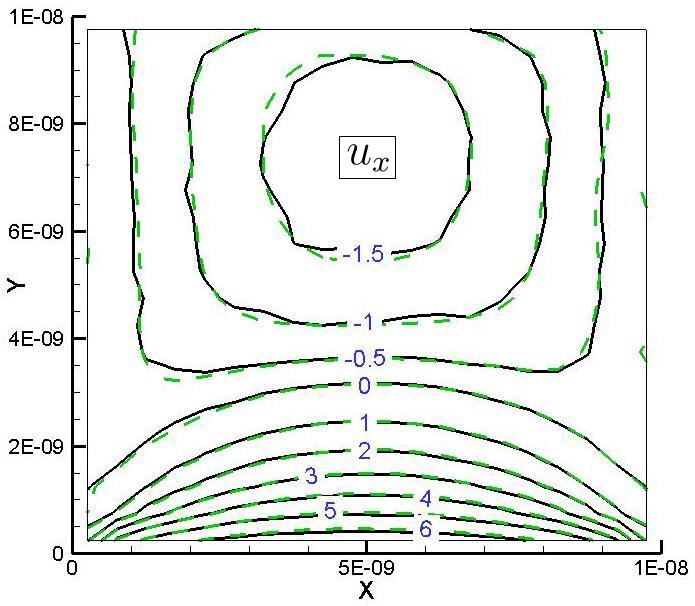}
  \includegraphics[width=0.45\textwidth]{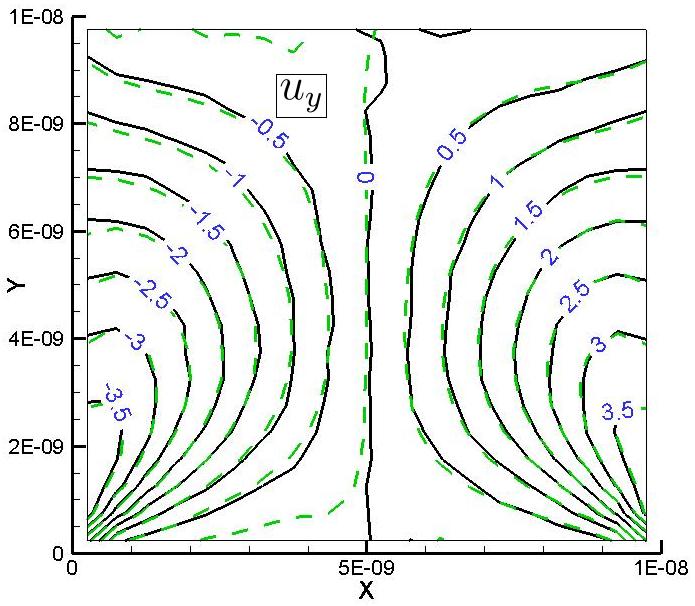}
  \caption{The steady state results of the lid-driven problem, $Kn=6.3$ and $U_\mathrm{wall}=20$ m/s; Dashed lines: DSBGK time-average results using 2000 simulated molecules per cell and 30 samples and totally taking about 9 minutes of CPU time, Solid lines: DSMC time-average results taking more than one day of CPU time.}
  \label{fig:CavityflowKn6.3U20-time-average}
\end{figure}

We choose the lid-driven problem of $Kn=6.3$ and $U_{\rm wall}=20$  m/s, where the density drift was observed in the previous DSBGK simulation which converges after about $40 \Delta t$ taking about 11 minutes of CPU time running on one processor of  Lenovo laptop E43A and has obvious deviation of density after $900 \Delta t$ (see Fig. 5 of \cite{Jun2012arXiv}), to verify the validity of the improved boundary condition and to show the elimination of the density drift. As in the previous DSBGK simulation, we use $\upsilon=\upsilon(\mu)$, the cell number is $20\times20$ and each cell contains about 2000 simulated molecules on average. The DSBGK simulation process converges also after about $40 \Delta t$, which takes about 5 minutes of CPU time running on one processor of MacBook Air. The difference of CPU time between the previous and current simulations is mostly due to using different laptops. We collect 30 samples sampled once at the end of each $\Delta t$ after the initial $40 \Delta t$. The total computational time is about 9 minutes of CPU time. Fig.~\ref{fig:CavityflowKn6.3U20-time-average} gives the validation of the DSBGK time-average results against the DSMC time-average results. The agreement between the DSBGK and DSMC methods in Fig.~\ref{fig:CavityflowKn6.3U20-time-average} is excellent and slightly better than that in Fig. 5 of \cite{Jun2012arXiv} where the agreement is very good though.

\begin{figure}[H]
\centering
\includegraphics[width=0.7\textwidth]{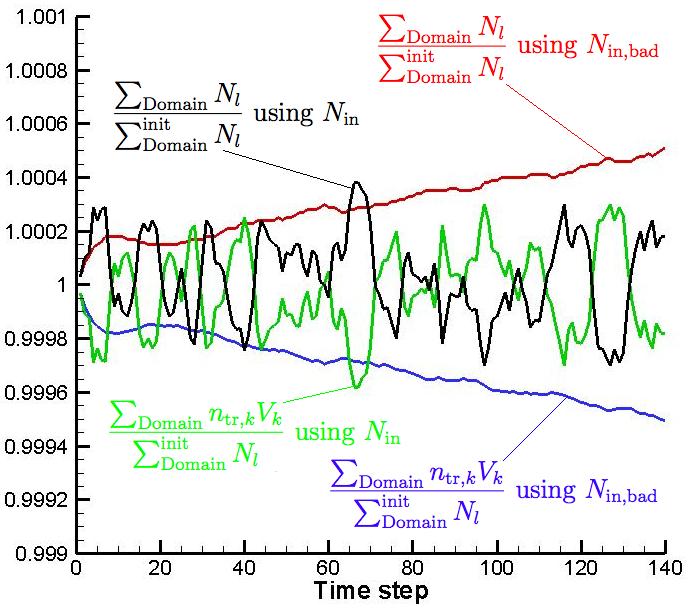}
\caption{Evolutions of the total numbers of real molecules represented by all simulated molecules and all cells, respectively, in the DSBGK simulations of the lid-driven problem using two different boundary conditions, $Kn=6.3$ and $U_\mathrm{wall}=20$ m/s, 2000 simulated molecules per cell.}
\label{fig:evolution of total number}
\end{figure}

We compute the transient total number of real molecules carried by all simulated molecules, namely $\sum_{\rm Domain} N_l$, and the total number represented by all cells, namely $\sum_{\rm Domain} n_{{\rm tr},k}V_k$.  The evolutions of this two quantities normalized by the initial total number $\sum^{\rm init}_{\rm Domain} N_l$ of real molecules is given in Fig.~\ref{fig:evolution of total number} which clearly shows that the unphysical density drift disappears after using the improved boundary condition. Thus, the efficient time-average process is valid for sampling all quantities in close as well as open problems when using the improved boundary condition. Fig.~\ref{fig:evolution of total number} also shows that the stochastic noise of the transient results is significantly increased when replacing the original boundary condition by the improved one which is more accurate but has more stochastic noise than the original one.

\section{Use the Original Diffuse Reflection Model in Problems with Tiny Perturbation}\label{s:simulations with original BCs}
\begin{figure}[H]
\centering
\includegraphics[width=0.45\textwidth]{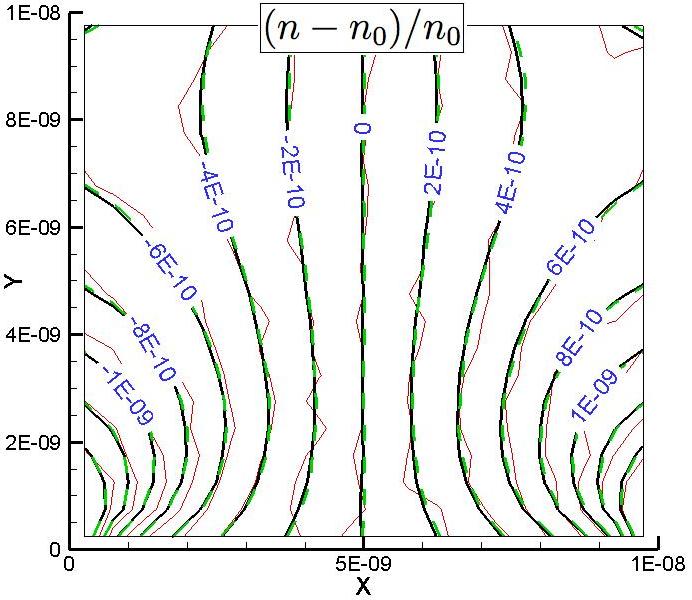}
\includegraphics[width=0.45\textwidth]{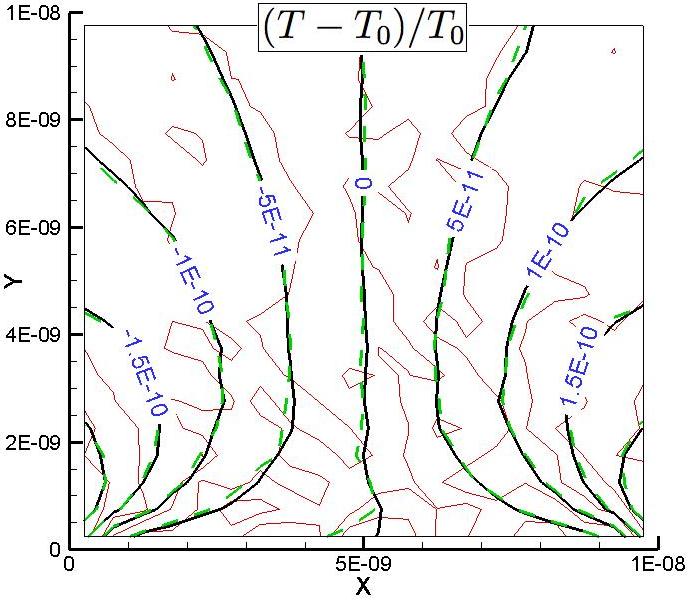} \\
\includegraphics[width=0.45\textwidth]{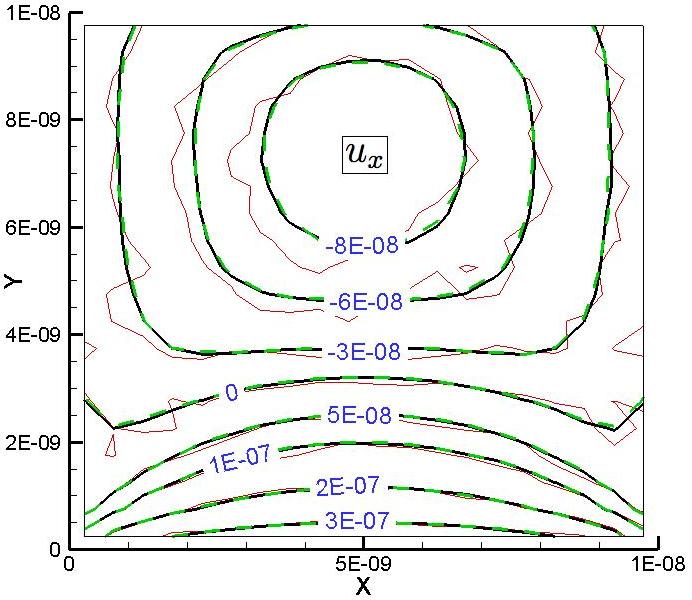}
\includegraphics[width=0.45\textwidth]{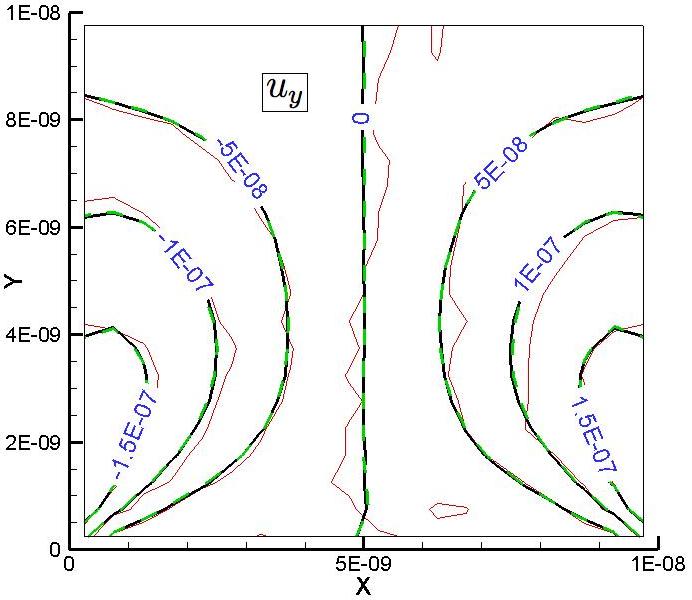}
\caption{The steady state results (without average) of the DSBGK simulations of the lid-driven problem using the \textit {original} boundary condition, $Kn=6.3$ and $U_\mathrm{wall}=10^{-6}$ m/s; Dashed green lines: results using 2000 simulated molecules per cell at the $40^{\rm th} \Delta t$ taking about 5 minutes of CPU time,  Solid black lines: results using 2000 simulated molecules per cell at the $900^{\rm th} \Delta t$, Thin solid red lines: results using 10 simulated molecules per cell at the $40^{\rm th} \Delta t$ (DSMC simulation theoretically requires $(\dfrac{20}{10^{-6}})^2$ days).} 
  \label{fig:CavityflowKn6.3-smallU}
\end{figure}

To show the high efficiency of the DSBGK method in simulating low-speed problems, we model the lid-driven problem again and only change the driven velocity $U_\mathrm{wall}$ from $20$ m/s to $10^{-6}$ m/s which is realistic in the MEMS or pore-scale flows of the shale gas where the pore size is about 10-100 nm. Unfortunately, the transient DSBGK results obtained using the improved boundary condition based on $N_{\rm in}$ are dominated by the stochastic noise. Thus, we suggest to use the original boundary condition based on $N_{\rm in,bad}$, which has much less noise than $N_{\rm in}$, because the difference between $N_{\rm in,bad}$ and the expected value of $N_{\rm in}$ is proportional (although may not linear) to the magnitude of perturbation which is tiny here. $n_{\rm eff,bad}$ computed using $N_{\rm in,bad}$ is almost exactly equal to the initial uniform number density $n_0$, which is physically accurate when the perturbation of the closed problem is tiny. In the DSBGK simulation using the original boundary condition, we use 2000 simulated molecules per cell again and thus the computational time for the initial $40 \Delta t$ is almost the same as in the previous case with $U_\mathrm{wall}=20$ m/s, namely about 5 minutes of CPU time. The simulation process converges also after about $40 \Delta t$ and the density drift at the $900^{\rm th} \Delta t$ is unperceivable as shown in Fig.~\ref{fig:CavityflowKn6.3-smallU}, where the DSBGK results at the $40^{\rm th} \Delta t$ obtained using only 10 simulated molecules per cell are given together to show the numerical stability of using few simulated molecules in closed problems. 

\section{Conclusions}\label{s:conc}
An improved diffuse boundary condition based on $N_{\rm in}$ is proposed to eliminate the density drift observed in the previous DSBGK simulation of \textit{closed} problem using the original boundary condition based on $N_{\rm in,bad}$. Consequently, the efficient time-average process is valid in the DSBGK simulations of closed as well as open problems. 

The unphysical density drift becomes unperceivable when the perturbation is tiny. Thus, for closed problems with tiny perturbation, we suggest to use the original boundary condition which makes the transient results smooth. In open problems, the original boundary condition is always an advisable choice to reduce the memory usage by using few simulated molecules. We don't need to resort to the improved boundary condition unless the density drift occurs and fails the goal of the simulations.
%

\end{document}